\documentclass[12pt]{iopart}

\usepackage{epsfig}

\newcommand {\be} {\begin{equation}}   
\newcommand {\ee} {\end{equation}}
\newcommand {\bea} {\begin{eqnarray}}
\newcommand {\eea} {\end{eqnarray}}
\newcommand {\bes} {\begin{displaymath}}
\newcommand {\ees} {\end{displaymath}}
\newcommand {\beas} {\begin{eqnarray*}}
\newcommand {\eeas} {\end{eqnarray*}}
\newcommand {\w} {\omega}

\begin{document}

\title{Chaperone assisted translocation}
\author{Tobias Ambj\"ornsson and Ralf Metzler}
\address{NORDITA (Nordic Institute for Theoretical Physics),\\ 
Blegdamsvej 17, 
DK-2100 Copenhagen \O,
Denmark.}
\ead{ambjorn@nordita.dk, metz@nordita.dk}

%\date{\today}

\begin{abstract}

We investigate the translocation of a stiff polymer through a nanopore
in a membrane, in the presence of binding particles (chaperones) that
bind reversibly to the polymer on both sides of the membrane. A bound
chaperone covers one (univalent binding) or many (multivalent binding)
binding sites. Assuming that the diffusion of the chaperones is fast
compared to the rate of translocation we describe the process by a
one-dimensional master equation. We expand previous models by a
detailed study of the effective force in the master equation, which is
obtained by the appropriate statistical mechanical average over the
chaperone states. The dependence of the force on the degree of valency
(the number of binding sites occupied by a chaperone) is studied in
detail. We obtain finite size corrections (to the thermodynamical
expression for the force), which, for univalent binding, can be
expressed analytically. We finally investigate the mean velocity for
translocation as a function of chaperone binding strength and
size. For both univalent and multivalent binding simple results are
obtained for the case of a sufficiently long translocating polymer.

\end{abstract}

\vspace{2cm}
Published in Physical Biology 1, 77 (2004).

\maketitle

\section{Introduction}

The problem of polymer translocation, i.e., the transport of an
oligomer or polymer (e.g., DNA, RNA or proteins) through a nanopore in
a membrane, is a process of fundamental importance in biology and
biotechnology. Relevant biological examples of this type of process
include: translocation of proteins through the endoplasmatic
reticulum, translocation of RNA through the nucleus pore membrane, the
viral injection of DNA into a host and DNA plasmid transport from cell
to cell through cell walls \cite{Alberts_etal}. Additionally,
biotechnological applications connected to membrane pore passaging,
such as rapid reading of DNA base sequences
\cite{Kasianowicz_etal2,Meller_Nivon_Branton}, analyte detection
\cite{Kasinowicz_Henrickson}, and nanosensor applications, have been
suggested. In medicine, controlled drug delivery is an ultimate goal,
a crucial element of which is the passage of cell and/or nuclei
membranes.

On the theoretical side there exist a number of investigations
\cite{Sung_Park,Lubensky_Nelson,Muthukumar,Muthukumar2,Chuang_Kantor_Kardar,Zandi_Reguera_etal,Flomenbom_Klafter,Ambjornsson_etal,Liebermeister,Elston,Tian_Smith,Di_Marzio_Kasianowicz,Kafri_Lubensky_Nelson,Meller},
whose common approach to the translocation problem is to employ a
one-dimensional description of the process using the penetration
length into the pore as a slow variable (``reaction'' coordinate);
each translocation step is assumed to be sufficiently slow so that the
polymer has time to relax to local equilibrium during the step (the
instantaneous relaxation approximation \cite{Boehm}). The dynamics is
then Markovian and can be described by a one-dimensional
Smoluchowski (Fokker-Planck) or master equation
\cite{Flomenbom_Klafter} in terms of the slow variable (however, as
pointed out in \cite{Chuang_Kantor_Kardar}, this approach breaks down
for very long polymers\footnote{For very long polymers, the process
becomes subdiffusive, and the Fokker-Planck equation may be replaced
by its fractional analogue \cite{Metzler_Klafter2}.}). The force
appearing in the Smoluchowski equation in general has entropic (chain
confinement in the pore reduces accessible degrees of freedom
\cite{Ambjornsson_etal}) as well as external (electric field,
chaperone binding etc.) contributions. Different theoretical studies
have focused on different experimentally measurable entities: The
mean translocation time is the most studied quantity
\cite{Sung_Park,Chuang_Kantor_Kardar}. More detailed studies
investigated the probability density of translocation times
\cite{Lubensky_Nelson,Flomenbom_Klafter}. Also the flux (number of
polymers passing through the pore per unit time) has been
theoretically investigated \cite{Ambjornsson_etal}. As pointed out already
there exist certain scenarios according to which the translocation
dynamics becomes subdiffusive
\cite{Chuang_Kantor_Kardar,Metzler_Klafter}. However in the present
work we concentrate on a system whose dynamics is Markovian.

Two important driving forces for translocation, both in vivo and in
experimental assays, are (i) an electric field across the membrane and
(ii) binding particles (chaperones). In this study the focus is on the
latter mechanism which appears to be particularly common for protein
translocation
\cite{Zandi_Reguera_etal,Liebermeister,Elston,Simon_Peskin_Oster,Rapoport,Neupert_Brunner},
but also of relevance for DNA transport through membranes
\cite{Salman_etal,Farkas_Derenyi_Vicsek}. In the careful investigation
by Simon et al. (see \cite{Simon_Peskin_Oster}) it was suggested that
the translocation of proteins is a simple thermal ratchet process,
i.e., the role of the chaperones is simply to prevent the backward
diffusion through the pore, thereby speeding up translocation. In
other studies the effect of the chaperones is modelled by using an
effective force originating from the chemical potential difference due
to the chaperones on the two sides of the membrane
\cite{Sung_Park,Lubensky_Nelson,Muthukumar,Muthukumar2,Kafri_Lubensky_Nelson}. More
recently \cite{Zandi_Reguera_etal}, the coupled
translocation-chaperone dynamics was investigated by Brownian
molecular dynamics simulation, initiating more detailed studies of the
chaperone-assisted translocation process. Although the studies above
provide insights into the role of the chaperones in the translocation
process, there is still no unified understanding.

In the present work we perform a detailed theoretical investigation of
the chaperone assisted translocation. The main contributions are (i)
our result for the finite size correction to the force; (ii) that we
include the possible occurrence of chaperones on both sides of the
pore; and (iii) that we consider also chaperones which are larger than
the size of a binding site. The paper is organized as follows: In
section \ref{sec:general_framework} we estimate the different relevant
timescales of the problem, and in particular we distinguish between
reversible and irreversible binding to the polymer. We also provide a
general framework in terms of a master equation, which allows a
theoretical description of the translocation dynamics. In section
\ref{sec:reversible_binding} we calculate the force on a stiff polymer
in the reversible binding regime by the appropriate statistical
mechanical average over the chaperone states. The polymer is divided
into $M$ equidistant binding sites to which the chaperones can bind,
and we distinguish between the cases when a bound chaperone covers one
(univalent binding) or several binding sites (multivalent
binding). For the case of univalent binding we recover previous
thermodynamical results for the force, however with a finite size
correction. In section \ref{sec:mean_velocity} we use the results from
the previous section in order to compute the mean velocity of the
polymer through the pore. For the case of sufficiently long stiff
polymers simple results are obtained for both univalent and
multivalent binding. In section \ref{sec:comparison} we compare the
effectiveness of the chaperone assisted translocation to electric
field driven translocation. Finally, in section
\ref{sec:summary_outlook} we give a summary and outlook.

\section{General framework and relevant timescales}\label{sec:general_framework}
In this section we provide a general framework for describing
sufficiently slow translocation dynamics in terms of a one-dimensional
master equation. By estimating relevant time-scales we distinguish
between three translocation regimes.

%%%%%%%%%%%%%%%%%%% Figure 1 %%%%%%%%%%%%%%%%%%%

\begin{figure}
  \begin{center}
    \scalebox{0.45}{\epsfbox{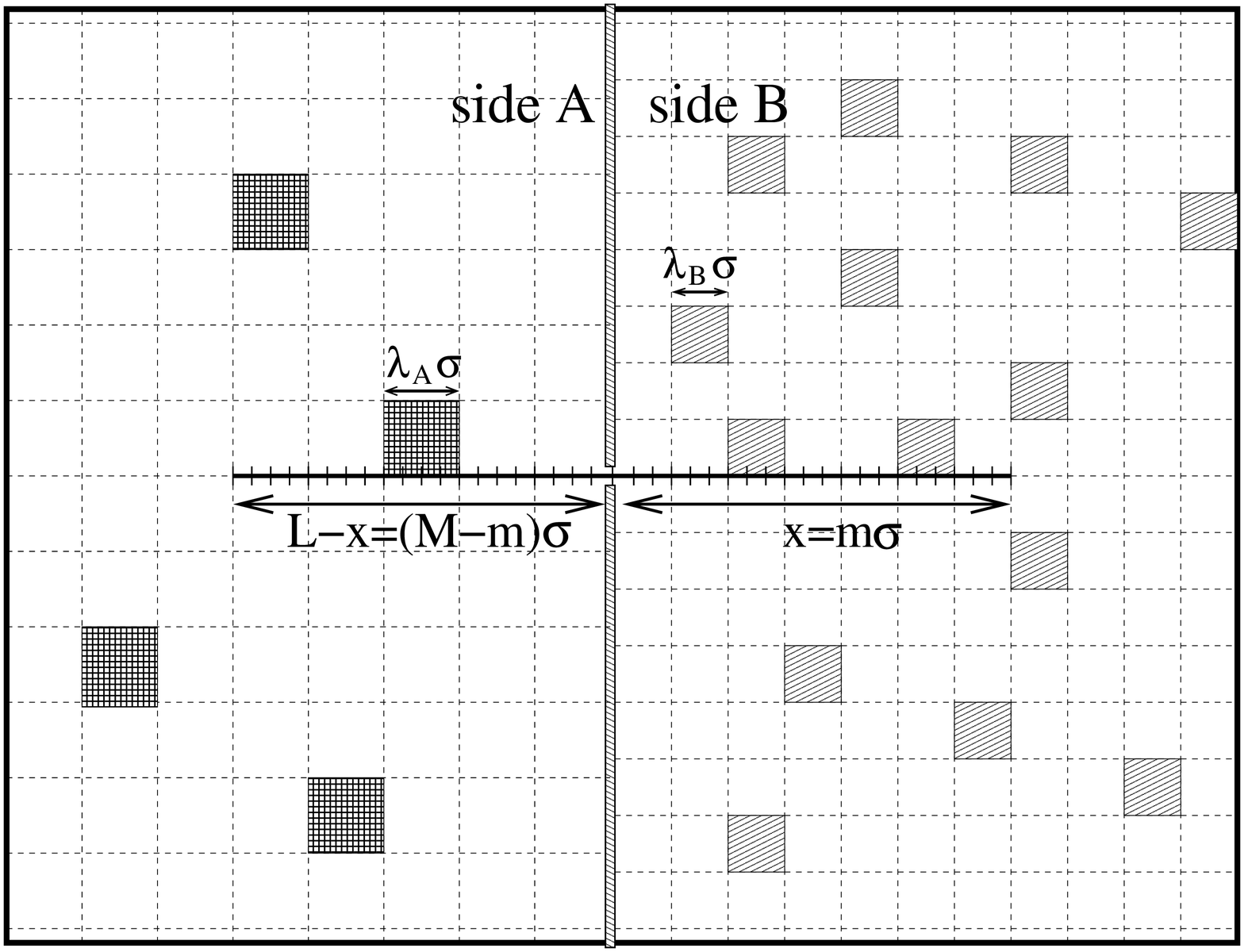}}
  \end{center}
  %\vspace{1cm}
  \caption{Translocation geometry in the presence of chaperones: A
  stiff polymer of length $L$ is translocating through a pore in a
  membrane. The filled boxes of the two sides are chaperones. Each
  chaperone on side $A$ (side $B$) occupies a volume $v_{0A}$ ($v_{0B}$),
  and binds to the polymer with a binding energy $\epsilon_A$
  ($\epsilon_B$). The volume of the compartment of side $A$ (side $B$) is
  $V_A$ ($V_B$). The size of a binding site is $\sigma$ and when a
  chaperone is attached to the polymer on side $A$ (side $B$) it occupies
  $\lambda_A$ ($\lambda_B$) binding sites. The total number of
  available binding sites is $M=L/\sigma$. The number of binding sites
  on side $B$ is $m=x/\sigma$, and the number of binding sites on side
  $A$ is $M-m=(L-x)/\sigma$.}
  \label{fig:geometry}
\end{figure}

%%%%%%%%%%%%%%%%%%%%%%%%%%%%%%%%%%%%%%%%%%%%%%%%%%%%%%%%%%%%

The geometry considered in this study is shown in figure\
\ref{fig:geometry}: A rod-like polymer is translocating through a
narrow pore in a membrane (the pore is usually a few nanometer in size,
corresponding to 10-15 nucleotides in the case of ssDNA and RNA
translocation through the $\alpha$-hemolysin channel). On each side of
the membrane there are chaperones which can bind to the polymer. We
here use the binding site size $\sigma$ (see figure
\ref{fig:geometry}) as the basic unit of length; for the case of
polynucleotides $\sigma$ may be the size of a base, and for unfolded
proteins $\sigma$ may correspond to the size of an aminoacid.  The
translocation process is then described by the variable $m$, which is
the number of binding sites on side $B$ (the distance the polymer has
entered into side $B$ is $x=m\sigma$). The total number of polymer
binding sites is $M=L/\sigma$ ($L$ is the length of the polymer) and
therefore for a given $m$ the number of binding sites on side $A$ is
$M-m$. Denote by $P(m,t)$ the probability that the polymer has passed
with $m$ binding sites into side $B$ at time $t$. We assume that
$P(m,t)$ satisfies a master equation:
 \footnote{A general master equation has the form \cite{Gardiner} $
\partial P(m,t)/\partial t=\sum_{m^\prime} \left( W(m|m^\prime )
P(m^\prime ,t)- W(m^\prime |m) P(m,t) \right) $, where $W(m|m^\prime
)$ are the transition probabilities per unit time. Assuming that
transitions can only occur in unit steps, i.e., $W(m |m^\prime )= t^+
(m^\prime)\delta_{m,m^\prime+1}+t^- (m^\prime)\delta_{m,m^\prime-1}$
we obtain equation \eref{eq:master_eq2}. }
 \bea
\frac{\partial}{\partial t}P(m,t)&=&t^+(m-1)P(m-1,t)+ t^-(m+1)P(m+1,t)\nonumber\\
& &-\left( t^+(m)+t^-(m)\right) P(m,t).\label{eq:master_eq2}
  \eea
\Eref{eq:master_eq2} is our starting point for studying the chaperone
assisted translocation process. Transitions can occur only one step in
the forward or backward direction (i.e, $m$ is increased or decreased
by one)\footnote{This assumption is reasonable in view of the fact
that the nanopore only allows a 1D array of monomers (aminoacids,
nucleotides) and that the passage is associated with a friction.}. The
transition probability for forward and backward motion is described by
the transfer coefficients $t^+(m)$ and $t^- (m)$, respectively.  In
order to have a complete description we must specify $t^+(m)$ and
$t^-(m)$ in terms of the fundamental parameters of the problem. We
choose these entities such that the Smoluchowski (Fokker-Planck)
equation is recovered in the limit $\sigma\rightarrow 0$.  This
requires
\footnote{In the continuum limit $\sigma\rightarrow 0$, equations
\eref{eq:master_eq2} and \eref{eq:t_m} satisfy the following
Smoluchowski (Fokker-Planck) equation \cite{Gardiner} $\partial
P(x,t)/\partial t= D \partial/\partial x \{-F(x)P(x,t)/k_B T+\partial
P(x,t)/\partial x \}$ where $x=m \sigma$ is the distance the polymer
has entered into side $B$. In the equation above the Einstein relation
$D=k_BT/\xi$ ($\xi$ is the friction constant for the polymer) is
implicit.}
  \bea
t^+(m)&=&\frac{1}{\tau_0} \left(1+\frac{F(m)}{2F_0}\right),\nonumber\\
t^-(m)&=&\frac{1}{\tau_0} \left(1-\frac{F(m)}{2F_0}\right),\label{eq:t_m}
  \eea
with a characteristic time $\tau_0=\sigma^2/D$ and characteristic
force $F_0=k_B T/\sigma$, where $D$ is the diffusion constant for the
polymer and $T$ is the temperature of the solvent ($k_B$ is the
Boltzmann constant). $F(m)$ is the force acting on the polymer; in
this work the focus is on obtaining relevant expressions for $F(m)$
when both sides of the membrane contain a certain population of
binding particles (chaperones), which can bind reversibly to the
polymer, as is illustrated in figure\ \ref{fig:geometry}. The
chaperones are larger than the pore size, so that there is no exchange
of chaperones between the two membrane sides. Typically the
translocation is driven by the binding of chaperones on the exit side
(side B). However, in vivo, chaperones are often present on both
membrane sides (possibly the role of the chaperones on the entrance
side is to unfold proteins before translocation
\cite{Simon_Peskin_Oster}), and we therefore allow for the presence of
chaperones in both compartments in figure \ref{fig:geometry}.  For
instance, protein import into mitochondria requires the presence of
chaperones both on the cytosolic side (proteins belonging to the
cytosolic hsp70 family) and on the mitochondrial side (mitochondrial
hsp70) \cite{Alberts_etal}. The use of a time-independent force in
equation \eref{eq:t_m} relies on the assumption that the chaperone
dynamics is fast compared to a characteristic translocation timescale
(see below). For a {\em flexible} polymer (not considered in detail in
this study) a time-independent force requires in addition that the
relaxation time of the polymer is small compared to relevant
translocation times.

Let us now investigate how fast a polymer translocates through the
pore, assuming, as in previous approaches, that once in the pore the
polymer is not allowed to fully retract to the entrance side. Keeping
in mind that the coordinate $m$ is then confined to the interval $0\le
m\le M$, and that we have a reflecting barrier at $m=0$, and an
absorbing state at $m=M+1$, the mean translocation time, for a process
described by the master equation \eref{eq:master_eq2}, becomes
\cite{Gardiner}
  \be
\tau=\tau_0 \sum_{m=0}^M \left(\Phi(m) \sum_{m^\prime =0}^m \frac{1}{t^+(m^\prime)\Phi (m^\prime)}\right)\label{eq:tau},
  \ee
with 
  \be
\Phi(m)=\prod_{u=1}^m \frac{t^-(u)}{t^+(u)}.
  \ee
We have above assumed that at time $t=0$ we start with an initial
condition at $m=0$.\footnote{In the continuum limit equation
\eref{eq:tau} becomes \cite{Gardiner} $\tau\simeq \int_{0}^L dx'' \exp(\beta
G(x''))\int_{0}^{x''}dx'\exp(-\beta G(x'))/D$ where $L$ is the
(contour) length of the polymer, $G(x)=-\int_{0}^xF(x')dx'$ and
$x=m\sigma$.}  For later reference we introduce an average velocity
  \be
\langle v \rangle \equiv \frac{L}{\tau}=v_0 \frac{\tau_0}{\tau} M\label{eq:v_mean},
  \ee
where we have defined a characteristic velocity $v_0\equiv
D/\sigma$. It is often more illustrative to use $\langle v
\rangle $ when discussing the results rather than the mean
translocation time $\tau$. For the case of a constant force $F(m)=F$
(so that the transition probabilities are independent of $m$,
$t^+(m)=t^+$ and $t^-(m)=t^-$) the mean translocation time can
straightforwardly be calculated using equations \eref{eq:t_m} and
\eref{eq:tau}, see \ref{sec:mean_translocation_time}.  The
corresponding mean velocity, for large $M$ (see
equation \eref{eq:tau_constant_force}), becomes
  \be
\langle v \rangle/v_0 =t^+-t^- = F/F_0\label{eq:v_mean_long}.
  \ee
Thus, for a constant force the mean velocity of a long polymer through
the pore is simply proportional to the force acting on the polymer,
i.e., the average motion is equivalent to classical motion, as it
should. As we will see in the subsequent section the force due to the
chaperones is such that it becomes constant for $m >m_0$, where $m_0$
is some characteristic finite size correction length. For a
sufficiently long polymer the finite size correction is negligible and
the expression above can be used to obtain the mean velocity. However,
very frequently the translocating chains are relatively short (for
instance, small proteins are $\sim 60$ aminoacids long, see reference
\cite{Alberts_etal} pp. 117-118; or in vitro studies of
single-stranded DNA translocation concern chain lengths well below 100
bases \cite{Meller_Nivon_Branton}), and the finite size corrections do
come into play.

There are three relevant time scales associated with the problem: The
time $\tau_{\rm diff}$ needed for the polymer to diffuse a distance of
the order of the binding site length $\sigma$; the typical time
$\tau_{\rm unocc}$ a binding site stays unoccupied; and the
characteristic time $\tau_{\rm occ}$ that a binding site remains
occupied. Let us estimate these different time scales, assuming that
there are no chaperones on the entrance side (side A) for
simplicity:\footnote{We do not consider the characteristic time
associated with one dimensional diffusion of binding proteins {\em
along} the polymer. For instance, for non-specific binding of DNA
transcription factors this introduces yet another time scale
\cite{Gerland_etal}.} The time needed for the translocating polymer to
diffuse a distance $\sigma$ is simply $\tau_{\rm diff}= \tau_0/2=
\sigma^2/2D$. Taking $\sigma\simeq $1 nm and $D\simeq$ 0.1
nm$^2$/s\footnote{We here use the estimated effective diffusion
constant $D\simeq$ 0.1 nm$^2$/s for protein import into mitochondria
in reference \cite{Chauwin_Oster_Glick}, obtained by comparison to
experimental translocation times. Notice that this value for the
diffusion constant is orders of magnitude smaller than that of a
freely diffusing polymer \cite{Chauwin_Oster_Glick}, probably due to
polymer interactions with the pore.} we find $\tau_{\rm diff}\simeq $
5 s. We now consider $\tau_{\rm unocc}$ and $\tau_{\rm occ}$. Denote
by $D_c$ the bulk diffusion constants for the chaperones and by $c_0$
the bulk concentration of chaperones. The chaperones bind to the
polymer site with a binding energy $\epsilon$ ($<$0). Clearly the
probability that a binding site is occupied depends on both the
concentration and the binding energy, and we will see in the next
section (see also \ref{sec:binding_partition_function}) that we can
form a dimensionless number $\kappa=c_0 K^{\rm eq}$ which is a
relevant measure of the binding strength, where we have defined an
equilibrium binding constant $K^{\rm eq}=v_0 \exp (\beta |\epsilon|)$
($v_0$ is the typical size of the chaperones, $\beta=1/(k_BT)$, with
$k_B$ being the Boltzmann constant and $T$ the temperature of the
solvent as before). Considering univalent binding for simplicity, the
equilibrium probability that a binding site is occupied is (see
\ref{sec:filling_fraction}) $P^{\rm eq}_{\rm occ}=\kappa/(1+\kappa)$,
and the probability that a binding site is unoccupied is therefore
$P^{\rm eq}_{\rm unocc}=1-P^{\rm eq}_{\rm occ}=1/(1+\kappa)$. At
equilibrium the ratio of these probabilities is equal to the ratio
between the times $\tau_{\rm occ}$ and $\tau_{\rm unocc}$, i.e. we
have $\tau_{\rm occ}/\tau_{\rm unocc}=\kappa$. We now proceed by
obtaining $\tau_{\rm unocc}$, which then through the above relation
also determines $\tau_{\rm occ}$: Consider a binding site, which
initially is vacant. If we assume that $\kappa$ is not too small, then
as soon as one chaperone is at the binding site it becomes trapped and
the binding site occupied. The distance between chaperones in solution
is $R\sim c_0^{-1/3}$. It suffices for a chaperone to diffuse a
distance of the order $R$ for any one chaperone to attach to the
binding site (provided $\kappa$ is sufficiently large), this takes a
time $\tau_{\rm unocc}\sim R^2/D_c\sim 1/(c^{2/3}D_c)$, which then
determines the characteristic time an initially vacant binding site
stays unoccupied (see reference \cite{Agmon_Gopich} for a more
thorough investigation of this problem).\footnote{Clearly, the
depletion of chaperones due to binding to a neighbouring binding site
could affect the time for binding. The concentrations of chaperones is
expected to be sufficiently high so that this effect will not be
dominant for estimating the relevant binding time.}  Taking
$D_c\simeq$ 10$^6$ nm$^2$/s and $c_0\simeq $10 $\mu$M, we find
$\tau_{\rm unocc}\simeq $ 1 ms. Thus typically the binding time is
faster than the time for the polymer to diffuse a distance of the
order one binding site. We note, however, that the above estimated
numerical values are very crude, and in particular that the polymer
diffusion constant $D$ may deviate substantially from the result given
here ($D$ depends on the nature of the polymer-pore interaction). We
found above that $\tau_{\rm occ}=\kappa\tau_{\rm unocc}$, and thus if
the binding strength $\kappa$ is large $\tau_{\rm occ}$ can become
large, even if $\tau_{\rm unocc}$ is small.  The considerations above
allow us to distinguish between three different dynamical regimes:\\
\noindent (i) {\em Diffusive regime}, $\tau_{\rm diff}\ll \tau_{\rm
unocc}, \tau_{\rm occ}$. In this regime the diffusion through the pore
is so fast that the chaperones do not have time to bind to the
translocating polymer. The force in equation \eref{eq:t_m} is then
essentially zero, and the mean translocation time equation
\eref{eq:tau} becomes simply $\tau=\tau_0 M^2/2$. This regime can
always be reached by lowering the concentration of chaperones
sufficiently. The diffusive regime corresponds to cases previously
discussed (see for instance \cite{Muthukumar,Sung_Park}) and will
therefore not be considered further in this investigation.\\
\noindent (ii) {\em Irreversible binding regime}, $\tau_{\rm unocc
}\ll \tau_{\rm diff}\ll \tau_{\rm occ}$. This regime corresponds to a
situation when the particles have sufficient time to bind, however do
not unbind from the polymer during the translocation. In this so
called Brownian ratchet regime
\cite{Simon_Peskin_Oster,Zandi_Reguera_etal,Sung_Park} the particles,
at sufficiently high concentrations and binding energies, bind
immediately as soon as the polymer has diffused a distance equal to
the size of a chaperone (or for univalent binding, a distance equal to
the distance between binding sites). The binding of chaperones
prohibits backward diffusion through the pore, and the polymer can
only ``jump'' in the forward direction. This regime can formally be
obtained by letting $F=2F_0$ in equation \eref{eq:t_m}; then the
backward transition probability is zero, $t^-=0$, the forward
transition probability per unit time is $t^+=2/\tau_0$, and hence the
master equation \eref{eq:master_eq2} becomes
  \be
\frac{\partial}{\partial t}P(m,t)=\frac{2}{\tau_0}\left( P(m-1,t)- P(m,t) \right).\label{eq:ratchet_eq}
  \ee
This equation gives a coarse-grained description of the ratchet
process, where only forward jumps are effectively allowed. The same
type of equation appears also in the theoretical description of shot
noise \cite{Gardiner}, and in the continuum limit corresponds to the
forward mode of the wave equation \cite{Landau_Lifshitz_ecm}. The mean
translocation time for ratchet motion becomes $\tau=\tau_0 M/2$ and
hence the mean velocity (see equation \eref{eq:v_mean}) is $\langle
v\rangle =2v_0$, which agrees with the result of other studies
\cite{Zandi_Reguera_etal,Simon_Peskin_Oster}. The ratchet mechanism
gives a decrease of the translocation time by a factor $\sigma/L=1/M$
compared to the translocation time in the diffusive regime described
in (i) above.\\
\noindent (iii) {\em Reversible binding regime}, $\tau_{\rm unocc
},\tau_{\rm occ}\ll \tau_{\rm diff}$. In this regime the particles
have time to bind and unbind many times during the time it takes for
the polymer to diffuse a distance $\sigma$. The polymer thus has time
to reach local equilibrium, and as we will see we can then obtain the
force $F(m)$ by the appropriate statistical mechanics average of the
chaperone states. We will in the rest of this paper investigate this
case of {\em reversible} (both univalent and multivalent) binding more
closely.

\section{Force $F(\lowercase{m})$ in the reversible binding regime}\label{sec:reversible_binding}

In this section we investigate the force $F(m)$ for reversible
binding of chaperones to the translocating polymer. The chaperones are
assumed to cover one binding site (univalent binding) or many binding sites
(multivalent binding), when attached to the polymer.

\subsection{General expression for the force $F(m)$}

We start by deriving a general expression for the force $F(m)$ on the
translocating polymer, arising from the interaction with chaperones on
the two sides of the membrane, in the reversible binding regime.

Let us obtain a statistical mechanical expression for the force on the
 translocating polymer. Denote by $Z(m,n_A,n_B)$ the
 Boltzmann-weighted number of configurations for a state specified by
 $m$, $n_A$ and $n_B$, where $n_A$ ($n_B$) is the number of attached
 chaperones on side $A$ (side $B$). For two unconnected compartments
 (see figure \ref{fig:geometry}) this statistical weight can be
 written as the product of the statistical weight on side $A$ and $B$
 respectively, i.e., $Z(m,n_A,n_B)=Z_A(m,n_A)Z_B(m,n_B)$. This is a
 natural decomposition, as the binding proteins cannot cross the
 nanopore (at least not in the presence of the translocating
 polymer). The force then decomposes in the form
  \be
F(m)=F_A(m)+F_B(m),
  \ee
i.e., the total force has independent contributions from side $A$ and
side $B$ respectively. Let us proceed by writing down the appropriate
statistical mechanical expression for the force originating from side
$\gamma$ ($\gamma$=$A$ or $B$). We have (see reference
\cite{Zandi_Reguera_etal})
  \bea
\frac{F_\gamma(m)}{F_0}&=&\sum_{n_\gamma=0}^{n_\gamma^{\rm max}} \frac{\beta \partial \ln Z_\gamma(m,n_\gamma)}{\partial m} \rho_\gamma^{\rm eq}(m,n_\gamma)\nonumber\\
&=&\frac{1}{Z_\gamma(m)}\sum_{n_\gamma=0}^{n_\gamma^{\rm max}} \frac{\partial }{\partial m}Z_\gamma(m,n_\gamma),\label{eq:F_gamma_def} 
  \eea
where $\rho_\gamma^{\rm eq}(m,n_\gamma)$ is the equilibrium
probability density for a state specified by $m$ and $n_\gamma$.  We
have above used the fact that the explicit expression for the
equilibrium distribution is $\rho_\gamma^{\rm
eq}(m,n_\gamma)=Z_\gamma(m,n_\gamma)/Z_\gamma(m)$, where the partition
function $Z_\gamma(m)$ for side $\gamma$ is obtained by summing the
statistical weight over all allowed values of $n_\gamma$:
  \be
Z_\gamma(m)=\sum_{n_\gamma=0}^{n_\gamma^{\rm max}} Z_\gamma(m,n_\gamma)\label{eq:Z_gamma},
  \ee
where $n_\gamma^{\rm max}$ is the maximum number of attached binding
particles on side $\gamma$. Notice that this quantity depends on
$m$. As before, we take $F_0=k_B T/\sigma$. The force, equation
\eref{eq:F_gamma_def}, is given by weighting the derivatives,
$-\partial \ln Z_\gamma(m,n_\gamma)/\partial m$, of the free energy of
state $m$, $n_\gamma$ by the equilibrium probability density (compare
to equation \eref{eq:n_exp_def}). We note here that in order for the
derivative in the force expression to be well-defined, the equation
for the free energy must be analytically continuable to non-integer
$m$ (the expression for the free energy considered here are
expressible in terms of factorials, which can be analytically
continued through $\Gamma$-functions, see next subsection). We point
out that the force $F_\gamma(m)$ may in general incorporate other
effects, for instance, as caused by electric fields or protein folding
(see discussions in sections \ref{sec:comparison} and
\ref{sec:summary_outlook}). If the polymer is flexible chain entropic
effects give an additional contribution to the force. Provided that
the chaperone binding is independent of the curvature of the polymer,
the binding force (as calculated in the next section) and the entropic
force are additive, and the entropic force expression given in
reference \cite{Sung_Park} can be used. For the sake of clarity, we
here neglect entropic effects, i.e., we assume a rod-like polymer.
This is not a strong restriction of the model, as in the presence of a
drift as exerted by the chaperones (see below), the entropic effect is
expected to be negligible for most chain lengths relevant to proteins,
compare reference \cite{Lubensky_Nelson}.

Let us finally rewrite the force in a form convenient for obtaining 
$F(m)$ in the thermodynamic limit ($m\rightarrow\infty$ for side $B$ and
$M-m\rightarrow\infty$ for side $A$). We write
  \be
F_\gamma(m)=\bar{F}_\gamma(m)-\triangle_\gamma(m)\label{eq:F_rewritten},
  \ee
with 
  \be
\frac{\bar{F}_\gamma(m)}{F_0} \equiv \frac{\partial Z_\gamma(m)/\partial m}{Z_\gamma(m)}=\frac{\partial }{\partial m}\ln Z_\gamma(m)\label{eq:F_thermodynamic},
  \ee
and 
  \be
\frac{\triangle_\gamma(m)}{F_0}\equiv\frac{\partial Z_\gamma (m) /\partial m - \sum_{n_\gamma=0}^{n_\gamma^{\rm max}} \partial Z_\gamma (m,n_\gamma) /\partial m}{Z_\gamma (m)}\label{eq:triangle},
  \ee
where the partition function $Z_\gamma(m)$ is given in equation
\eref{eq:Z_gamma}. The force \eref{eq:F_rewritten} is composed of two
terms. The first term, explicitly given by equation
\eref{eq:F_thermodynamic}, is the thermodynamic expression for the
force obtained by taking the derivative of the logarithm of the
partition function with respect to $m$. As we will see in subsections
\ref{sec:univalent_binding} and \ref{sec:general_results_large_m} this
term is in general independent of $m$, and is proportional to the
chemical potential difference across the membrane. The second term,
given in equation \eref{eq:triangle}, is a correction term to this
thermodynamic result. We note that the presence of this correction
term is due to the fact that the upper limit $n_\gamma^{\rm max}$ in
the sum in \eref{eq:triangle} depends on $m$ (if this were not the
case we could move the derivative in front of the sum and
$\triangle_\gamma(m)$ would be identically zero). In subsection
\ref{sec:univalent_binding} we show that $\triangle_\gamma(m)$
vanishes for large $m$ for the case of univalent binding. We will
therefore henceforth call $\triangle_\gamma(m)$ a finite size
correction term. We note here that in order to calculate the general
force expression equation \eref{eq:F_gamma_def}, we must evaluate
(complicated) sums involving the statistical weights
$Z_\gamma(m,n_\gamma)$. However, for obtaining the force equation
\eref{eq:F_thermodynamic} in the thermodynamic limit (i.e., large
protrusion distances), a knowledge of the partition function $Z_\gamma
(m)$ suffices (as we will see in subsection
\ref{sec:general_results_large_m}, $Z_\gamma(m)$ can be
straightforwardly calculated using a transfer matrix
approach).

\subsection{Explicit expression for the force $F(m)$}

In this subsection we study in more detail the forces $F_A(m)$ and
$F_B(m)$ when the two compartments contain chaperones, which bind
reversibly to the translocating polymer. In particular, we obtain the
forces as a function of chaperone effective binding strengths and sizes.

In order to obtain the force on the polymer we must have explicit
expression for the statistical weights $Z_\gamma(m,n_\gamma)$ (see
equation \eref{eq:F_gamma_def}; $\gamma$=$A$ or $B$). The details of
the calculation of $Z_\gamma(m,n_\gamma)$ are given in
\ref{sec:binding_partition_function}. There are two entropic effects
that must be taken into account: (i) as $m$ increases the number of
available binding sites increases on side $B$ and vice versa; and (ii)
as the number $n_\gamma$ of bound chaperones increases the entropy of
the surrounding ``gas'' decreases. We neglect the reduction of volume
due to the presence of the translocating polymer. We assume that the
chaperones are equal in size (univalent binding) or larger than
(multivalent binding) the size of a binding site, and cover an integer
$\lambda_\gamma$ ($\ge 1$) number of binding sites if bound to the
polymer on side $\gamma$ (for instance, bacterial transcription
factors cover $\sim 10-20$ basepairs \cite{Gerland_etal}). The
statistical weights then become (see equation \eref{eq:Z_B_spec_app})
  \be
Z_\gamma(m,n_\gamma) = \Omega_\gamma^{\rm bind}(m,n_\gamma) \kappa_\gamma^{n_\gamma}\label{eq:Z_B_spec},
  \ee
where $\Omega_\gamma^{\rm bind}(m,n_\gamma)$ denotes the number of
ways of arranging $n_\gamma$ particles onto the $m$ binding sites on
side $B$ or onto the $M-m$ binding sites on side $A$. For the case of
dilute solutions the effective binding strength $\kappa_\gamma $
appearing above can be written:
  \be
\kappa_\gamma = c_\gamma K^{\rm eq}_\gamma  \label{eq:kappa_gamma}
  \ee
where $c_\gamma $ is the concentration of chaperones on side $\gamma$,
and $K^{\rm eq}_\gamma$ is the equilibrium binding constant, see
equation \eref{eq:eq_binding_constant}. Since the effective binding
strength $\kappa_\gamma$ is proportional to the chaperone
concentration, one can experimentally vary $\kappa_\gamma$ by changing
the latter. We note that for $\lambda_\gamma \ge 2$ there exist
correlations between binding sites in the sense if one binding site is
occupied, then at least one of the neighbouring binding sites is
occupied. In turn, a binding protein needing more than one binding
site to actually bind cannot bind between already bound chaperones if
their distance is less than $\lambda_\gamma$. Thus, the binding
characteristics for large and small ($\lambda_\gamma=1$) are
different. Explicitly we have for side $B$ \cite{Epstein,McQuistan}
  \be
 \Omega_B^{\rm bind}(m,n_B)={m-(\lambda_B-1)n_B \choose n_B}=\frac{(m-(\lambda_B-1)n_B)!}{n_B! (m-\lambda_B n_B)!}\label{eq:Omega_B_nonspec},
  \ee
and similarly for side $A$ with the replacement $m\rightarrow M-m$,
$n_B\rightarrow n_A$ and $\lambda_B\rightarrow \lambda_A$.  We note
here that we have above neglected cooperative effects (i.e., the
effect that the chaperones attached to the polymer may interact). Such
effects are usually incorporated into the theory through a
cooperativity parameter $\w_\gamma$
\cite{Epstein,DiCera_Kong,McGhee_vonHippel}, where no cooperativity
corresponds to $\w_\gamma=1$. For $\w_\gamma>1$ (positive
cooperativity) the chaperones bound to the polymer interact
attractively, whereas for $\w_\gamma <1$ (negative cooperativity) the
chaperones repel each other.  The (somewhat lengthy) relevant
expressions for $Z_\gamma (m,n_\gamma)$ with $\w_\gamma\neq 1$ can be
found in reference \cite{Epstein}.  In subsection
\ref{sec:general_results_large_m}, we revisit the problem and show
that the partition function $Z_\gamma (m)$ for large $m$ may be
straightforwardly obtained including cooperativity, allowing for a
determination of the force $F(m)$ (see equation
\eref{eq:F_thermodynamic}) in the thermodynamic limit.

Equations \eref{eq:Z_B_spec}, \eref{eq:kappa_gamma} and
\eref{eq:Omega_B_nonspec} for the statistical weight $Z_\gamma
(m,n_\gamma)$ completely determine the effective force $F_\gamma(m)$
(see equation \eref{eq:F_gamma_def}). Combining equations
\eref{eq:F_gamma_def} and \eref{eq:Z_B_spec} we straightforwardly
obtain the force on the polymer from side $A$:
  \be
\fl \frac{F_A(m)}{F_0}=-\sum_{n_A=0}^{n_A^{\rm max}}\rho^{\rm eq}_A(m,n_A)\{\Psi(M-m-(\lambda_A -1)n_A +1)-\Psi(M-m-\lambda_A n_A +1)\}\label{eq:F_A_general}.
  \ee
Similarly the force from side $B$ is:
  \be
\fl \frac{F_B(m)}{F_0}=\sum_{n_B=0}^{n_B^{\rm max}}\rho^{\rm eq}_B(m,n_B)\{\Psi(m-(\lambda_B -1)n_B +1)-\Psi(m-\lambda_B n_B +1)\}\label{eq:F_B_general},
  \ee
where $\Psi(z)=d \ln \Gamma (z)/dz=\Gamma'(z)/\Gamma(z)$ is the
$\Psi$-function \cite{Abramowitz_Stegun}, and we have analytically
continued the factorials appearing in equation
\eref{eq:Omega_B_nonspec} using $\Gamma$-functions. The maximum number of
particles that can attach to the polymer on side $B$ is $n_B^{\rm
max}=[m/\lambda_B]$, i.e., it is the largest integer smaller than or
equal to $m/\lambda_B$. Similarly for side $A$ the maximum number of
attached chaperones is $n_A^{\rm max}=[(M-m)/\lambda_A]$. We notice that the
force from side $B$ is zero, as it should, when the chain does not
protrude at that side, i.e., we have $F_B(m=0)=0$. Also, since
$\Psi(z)$ is an increasing function with $z$, the force from side $B$
is positive $F_B(m)\ge 0$, whereas the force from side $A$ is negative
$F_A(m)\le 0$. Equations \eref{eq:F_A_general} and
\eref{eq:F_B_general} are general expressions for the force, and are
convenient for numerical computations. We point out that in general
the force $F(m)$ depends on five dimensionless variables: the binding
strengths $\kappa_A$ and $\kappa_B$, the relative sizes of the
chaperones $\lambda_A$ and $\lambda_B$, as well as the effective
length $M=L/\sigma$ of the polymer. In the next two subsections we
derive simplified approximate results for the cases: (i) univalent
binding ($\lambda_\gamma=1$), and (ii) general multivalent binding and
long polymers. In the latter subsection we also revisit the problem of
cooperative effects ($\omega_\gamma\neq 1$). In these two subsections
we compare the results to the exact expressions for the force,
equations \eref{eq:F_A_general} and \eref{eq:F_B_general}.

\subsection{Univalent binding}\label{sec:univalent_binding}

In this subsection we consider the case of univalent binding,
  $\lambda_\gamma=1$, and obtain the force in the thermodynamic
  limit. We also derive a an approximate expression for the finite
  size correction to the force.
 
For univalent binding $\lambda_\gamma=1$ we have $n_B^{\rm
  max}=m$ and $n_A^{\rm max}=M-m$ and therefore $\Omega_\gamma^{\rm
  bind}(m,n_\gamma)=n_\gamma^{\rm max}!/(n_\gamma!  (n_\gamma^{\rm
  max}-n_\gamma)!)$ (see equation \eref{eq:Omega_B_nonspec}). The
  partition functions $Z_A(m)$ and $Z_B(m)$ can straightforwardly be
  calculated using equation \eref{eq:Z_B_spec}. We find
  \be
Z_\gamma(m)=\sum_{n_\gamma =0}^{n_\gamma^{\rm max}}Z_\gamma(m,n_\gamma)=(1+\kappa_\gamma)^{n_\gamma^{\rm max}},\label{eq:Z_x_spec}
  \ee
where we have used the binomial theorem \cite{Abramowitz_Stegun}.
This equation is a standard result for the partition function for
univalent, non-cooperative binding to a polymer
\cite{Wyman_Gill}. Notice that the dependence on the chaperone
concentration and the binding energies appear only through the
quantity $\kappa_\gamma$ (see equation \eref{eq:kappa_gamma}). Let us
now calculate the force $\bar{F}_\gamma(m)$ in the thermodynamic
limit. Using equations \eref{eq:F_thermodynamic} and
\eref{eq:Z_x_spec} we find:
  \be
\frac{\bar{F}_\gamma(m)}{F_0}=\frac{\bar{F}_\gamma}{F_0}=\pm \ln (1+\kappa_\gamma)\label{eq:g_B}, 
  \ee
where the plus sign corresponds to side $B$, and the minus sign
corresponds to side $A$. We notice that the thermodynamic force is
independent of $m$, and the expression above can be viewed as a
chemical potential difference across the membrane, compare references 
\cite{Zandi_Reguera_etal,Kafri_Lubensky_Nelson,Farkas_Derenyi_Vicsek}.
 
\begin{figure}
  \begin{center}
    \scalebox{0.8}{\epsfbox{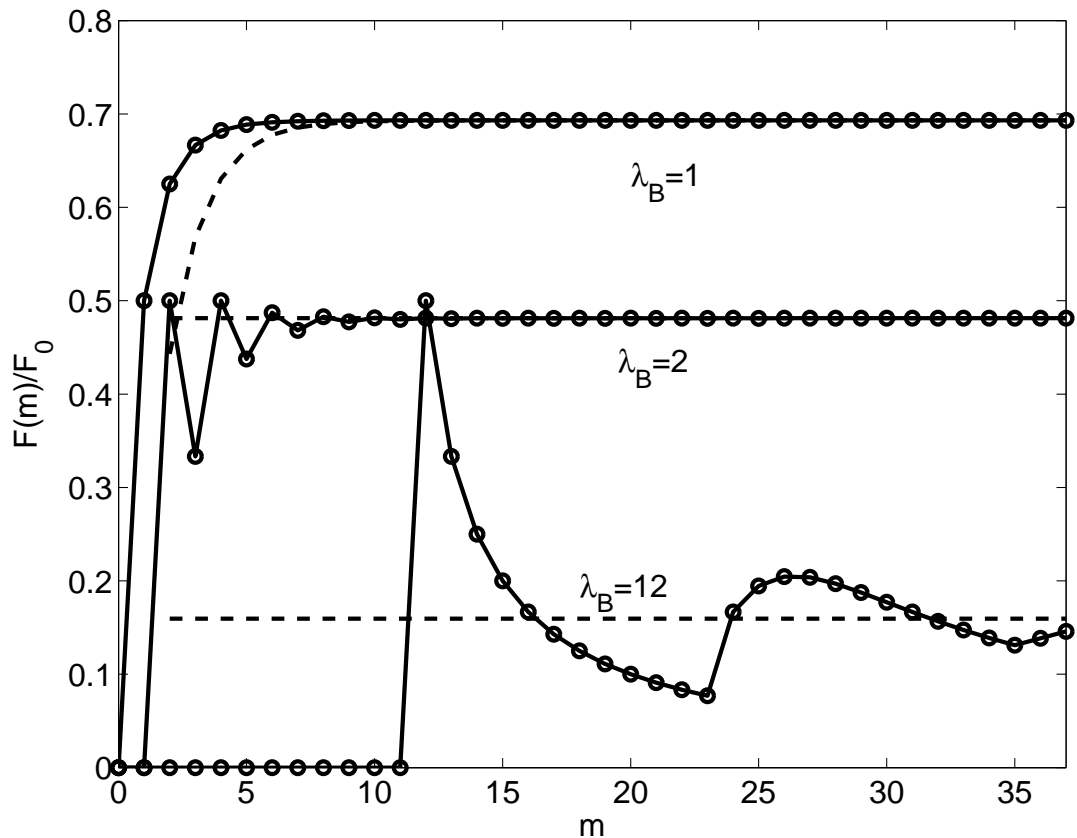}}
  \end{center}
  %\vspace{1cm}
  \caption{The force $F(m)/F_0$ in units of $F_0=k_B T/\sigma$ ($T$ is
  the temperature of the solvent, $k_B$ is the Boltzmann constant and
  $\sigma$ is the binding site size, see figure
  \ref{fig:geometry}). The dashed lines correspond to approximate
  results: For the cases of univalent binding $\lambda_B=1$ the upper
  dashed line is the analytic result as contained in equation
  \eref{eq:g_B}. For the case $\lambda_B=2$ we have included a plot
  (middle dashed line) of the approximate result contained in equation
  \eref{eq:F_dimer}. For $\lambda_B=12$, a typical value for many
  DNA-binding proteins, the lower dashed line corresponds to the
  result obtained through equations \eref{eq:algebraic_eq} and
  \eref{eq:F_DiCera}.  The remaining curves corresponds to the results
  obtained through the exact expression equation
  \eref{eq:F_B_general}, for: $\lambda_B=1$ (upper curve),
  $\lambda_B=2$ (middle curve) and $\lambda_B=12$ (lower curve).  The
  binding strength (see equation \eref{eq:kappa_gamma}) was taken to
  be $\kappa_B=1$. No binding particles were assumed to be present on
  side $A$. Notice that for small $m$ (and $\lambda_B\ge 2$) the force
  ``oscillates'' with a period $\lambda_B$. For large $m$ the force
  approaches a constant value. The onset of the force is where
  $m=\lambda_B$, i.e., the force is zero unless there are sufficiently
  many binding sites on side $B$ to accommodate at least one
  chaperone. The solid lines are only meant to guide the eye. For
  typical chaperone sizes and translocating polymer lengths the finite
  size corrections may become quite relevant. }
  \label{fig:v_x}
\end{figure}

We proceed by considering the finite size correction to the force,
equation \eref{eq:triangle}. Replacing the sum over $n_\gamma$ in
equation \eref{eq:triangle} by an integration and using Leibniz's
theorem for differentiation of an integral \cite{Abramowitz_Stegun} as
well as the fact that $Z(m,n_\gamma^{\rm max})=\kappa_\gamma^{
n_\gamma^{\rm max}}$ we find (for side $B$)
  \be
\frac{\triangle_B(m)}{F_0}\approx \left( \frac{\partial n_B^{\rm max}}{\partial m}\right) \frac{Z_B(m,n_B^{\rm max})}{Z_B(m)}=\left( \frac{\kappa_B}{1+\kappa_B}\right)^m=\exp(-m/m_{0B})\label{eq:triangle2}, 
  \ee
where $m_{0B}=1/\ln ([1+\kappa_B]/\kappa_B)=1/\ln (f_B^{-1})$ in terms
of the filling fraction $f_B$ of the polymer on side $B$ as contained
in equation \eref{eq:f_gamma}. The finite size correction for side
$A$, $\triangle_A(m)$, is obtained in an identical fashion. The finite
size correction is exponentially decreasing with increasing $m$, and
vanishes over distances larger than $m_{0\gamma}$, where $m_{0\gamma}$
is determined by the filling fractions on the two sides; for large
(small) filling fraction, i.e., large (small) $\kappa_\gamma$, the
correction decays slowly (rapidly) with $m$. Figure \ref{fig:v_x}
shows the above expression for the force (equations
\eref{eq:F_rewritten}, \eref{eq:g_B} and \eref{eq:triangle2}) together
with the exact result (equation \eref{eq:F_B_general}). Notice that
the result above, i.e. the expression the force as given by equations
\eref{eq:F_rewritten}, \eref{eq:g_B} and \eref{eq:triangle2}, captures
the decrease of the force with decreasing $m$, but does not fully
agree with the result from the general expression for the force
\eref{eq:F_B_general}, due to the continuum approximation leading to
equation \eref{eq:triangle2}. We point out that a decrease of $F_B(m)$
for decreasing $m$ agrees with the molecular dynamics simulations in
reference \cite{Zandi_Reguera_etal}. The results obtained in this
subsection show that for a long ($m>m_{0B}$) polymer the force can be
calculated using the thermodynamic expression
$\bar{F}_\gamma(m)/F_0=\partial \ln Z_\gamma(m)/\partial m$, but for
short polymers finite size corrections become relevant, and one has to
resort to the exact expressions for the force, equations
\eref{eq:F_A_general} and \eref{eq:F_B_general}. In figure
\ref{fig:v_x}, this fact is demonstrated for the case $\lambda_B=12$, a
typical value for binding proteins.

%%%%%%%%%%%%%%%%%% Figure 2 %%%%%%%%%%%%%%%%%%%%%%%%%%%%%%%%

%%%%%%%%%%%%%%%%%%%%%%%%%%%%%%%%%%%%%%%%%%%%%%%%%%%%%%%%%%%%%%%%

\subsection{General case, large $m$}\label{sec:general_results_large_m}

In this subsection we show that for large $m$ the force
$F_\gamma(m)$ can be obtained through the solution of an algebraic
equation for general $\lambda_\gamma$. The approach allows us to
revisit the problem of cooperativity ($\w_\gamma\neq 1$) in a
straightforward manner.

In order to obtain the force for large $m$, a knowledge of the
partition function $Z_\gamma(m)$ suffices (see
equation \eref{eq:F_thermodynamic}). Rather than using the combinatorial
approach given in the previous subsections the partition function can
more conveniently be obtained using the approach pursued in reference
\cite{DiCera_Kong} (see also \cite{Kong}): In general the partition
function $Z_\gamma(m)$ can be written as \cite{DiCera_Kong} 
  \be
Z_\gamma(m)= \sum_{j=1}^{\lambda_\gamma+1}\alpha_j \Lambda_j^m\label{eq:Z_DiCera},
  \ee
where $\Lambda_j$ are the $\lambda+1$ roots to the algebraic equation
($\gamma$=$A$ or $B$)
  \be
\Lambda^{\lambda_\gamma+1}-\Lambda^{\lambda_\gamma}-\omega_\gamma \kappa_\gamma \Lambda +(\omega_\gamma-1)\kappa_\gamma=0.\label{eq:algebraic_eq}
  \ee
The prefactors $\alpha_j$ are independent of $m$ and are explicitly
given by $\alpha_j=\Lambda_j-\lambda_\gamma d\Lambda_j/d \ln
\kappa_\gamma$. \Eref{eq:algebraic_eq} is the secular equation
associated with the transfer matrix of the system
\cite{DiCera_Kong}. Equations \eref{eq:Z_DiCera} and
\eref{eq:algebraic_eq} completely determine the partition function of
the system. Notice that the above approach incorporates cooperativity
effects (through the cooperativity parameter $\omega_\gamma$) without
substantially raising the level of complexity. However, we point
out that in the present approach only the partition function
$Z_\gamma(m)$, and not the statistical weights $Z_\gamma(m,n_\gamma)$,
can be calculated. Therefore the approach discussed in this subsection
does not allow computation of the exact expression for the force (see
the definition of the force, equation \eref{eq:F_gamma_def}, and also
equations \eref{eq:F_A_general} and \eref{eq:F_B_general}).

Let us now calculate the force for large $m$. Denote by
$\Lambda_\gamma^{\rm max}$ the largest root to the algebraic equation
equation \eref{eq:algebraic_eq}. Then for large $m$ the force equation
\eref{eq:F_thermodynamic} becomes
  \be
\frac{\bar{F}_\gamma(m)}{F_0}=\frac{\bar{F}_\gamma}{F_0}\approx \pm \ln \Lambda_\gamma^{\rm max}\label{eq:F_DiCera},
  \ee
where the plus sign corresponds to $\gamma=B$, and the minus sign to
$\gamma=A$.  The force in the thermodynamic limit is hence
proportional to the logarithm of the largest root $\Lambda_\gamma^{\rm
max}$. \Eref{eq:algebraic_eq} can straightforwardly be solved on a
computer, and hence the determination of the force for large $m$ is a
simple matter. For univalent, non-cooperative ($\w_\gamma=1$)
binding equation \eref{eq:algebraic_eq} becomes a first order
algebraic equation which can be easily solved. The force as obtained
from this solution together with equation \eref{eq:F_DiCera} agrees
with equation \eref{eq:g_B} as it should. In the previous subsection
we showed that the force in the thermodynamic limit is independent of
$m$ for the case of univalent, non-cooperative binding. The result
above proves that the force, for large $m$, is independent on $m$ for
general values of $\lambda_\gamma$ and $\w_\gamma$.  We have in figure
\ref{fig:v_x} plotted the force for multivalent binding using the
exact result, equation \eref{eq:F_B_general}, as well as the above
result $\bar{F}_B(m)$.  The
agreement is good for large $m$. Notice that the exact result has an
``oscillatory'' behaviour with a period $\lambda_B$ for small
$m$-values. We interpret the these oscillations in the following way:
if $m$ is equal to an integer multiple of $\lambda_B$ the polymer can,
potentially (for large binding strengths), fill the polymer and hence
completely restrict backward motion (perfect 'ratcheting'). However,
when $m$ is not an integer multiple of $\lambda_B$ there must be
vacant spaces in between bound chaperones (for instance for $m=5$ the
maximum number of bound chaperones is 2 for divalent binding
($\lambda_B=2)$, and hence there must be at least one vacant binding
site, even for large binding strengths), and the 'ratchet' effect is
less pronounced. In reference \cite{Epstein} similar types of
oscillations were found in the filling fraction of a polymer as a
function of $m$, for the case of multivalent binding.

For divalent ($\lambda_\gamma =2$), non-cooperative binding
equation \eref{eq:algebraic_eq} becomes a second order algebraic
equation, which can straightforwardly be analytically solved, yielding
$\Lambda=[1\pm (1+4\kappa_\gamma)^{1/2}]/2$. The corresponding force
(for large $m$) becomes
  \be
\frac{\bar{F}_\gamma(m)}{F_0}=\frac{\bar{F}_\gamma}{F_0}\approx \pm \ln \left( \frac{1+(1+4\kappa_\gamma)^{1/2}}{2}\right)\label{eq:F_dimer},
  \ee
where the plus sign corresponds to $\gamma=B$, and the minus sign to
$\gamma=A$. We point out that the force for divalent
binding as given by equation \eref{eq:F_dimer} has a different
functional dependence on $\kappa_\gamma$ compared to the case of
univalent binding, see equation \eref{eq:g_B}.

Let us finally obtain the force for large $m$, univalent binding
($\lambda_\gamma=1$) and {\em including} cooperativity effects
 (arbitrary $\omega_\gamma$) using the approach above. For this case
equation \eref{eq:algebraic_eq} becomes a second order algebraic equation
with the roots: $\Lambda=(1+\omega_\gamma\kappa_\gamma)/2\pm
\{(1+\omega_\gamma\kappa
)^2/4+(\omega_\gamma-1)\kappa_\gamma\}^{1/2}$. Hence the force,
equation \eref{eq:F_DiCera}, becomes 
  \be
\frac{\bar{F}_\gamma}{F_0}\approx \pm \ln \left\{ \frac{1+\omega_\gamma \kappa_\gamma}{2}+\left( \left( \frac{1+\omega_\gamma \kappa_\gamma}{2}\right)^2+\left(\omega_\gamma-1\right)\kappa_\gamma \right)^{1/2} \right\}\label{eq:F_univalent_coop},
  \ee
where we have assumed that $\omega_\gamma \ge W (\kappa_\gamma)\equiv
3[2(2+\kappa_\gamma)^{1/2}/3-1]/\kappa_\gamma$ so that the roots are
real. We note that $W(\kappa_\gamma)$ has a maximum value $1/2$; therefore
equation \eref{eq:F_univalent_coop} applies whenever the cooperativity
parameter satisfies $\w_\gamma \ge 1/2$. In addition, when
$\kappa_\gamma < 1/4$ we find that $W(\kappa_\gamma)$ is
negative. Hence equation \eref{eq:F_univalent_coop} is valid for any
value of the cooperativity parameter $\w_\gamma$ provided that
$\kappa_\gamma\le 1/4$.  For no cooperativity $\omega_\gamma=1$ the
above result reduces to previous results (see equation
\eref{eq:g_B}). Notice that the force for the case of positive
cooperativity is larger than the force for negative cooperative
binding, as it should.

\section{Mean velocity}\label{sec:mean_velocity} 

In this section we study the mean velocity for the polymer
translocation. In particular we find a simple form for the mean
velocity for long polymers.

\begin{figure}
  \begin{center}
    \scalebox{0.8}{\epsfbox{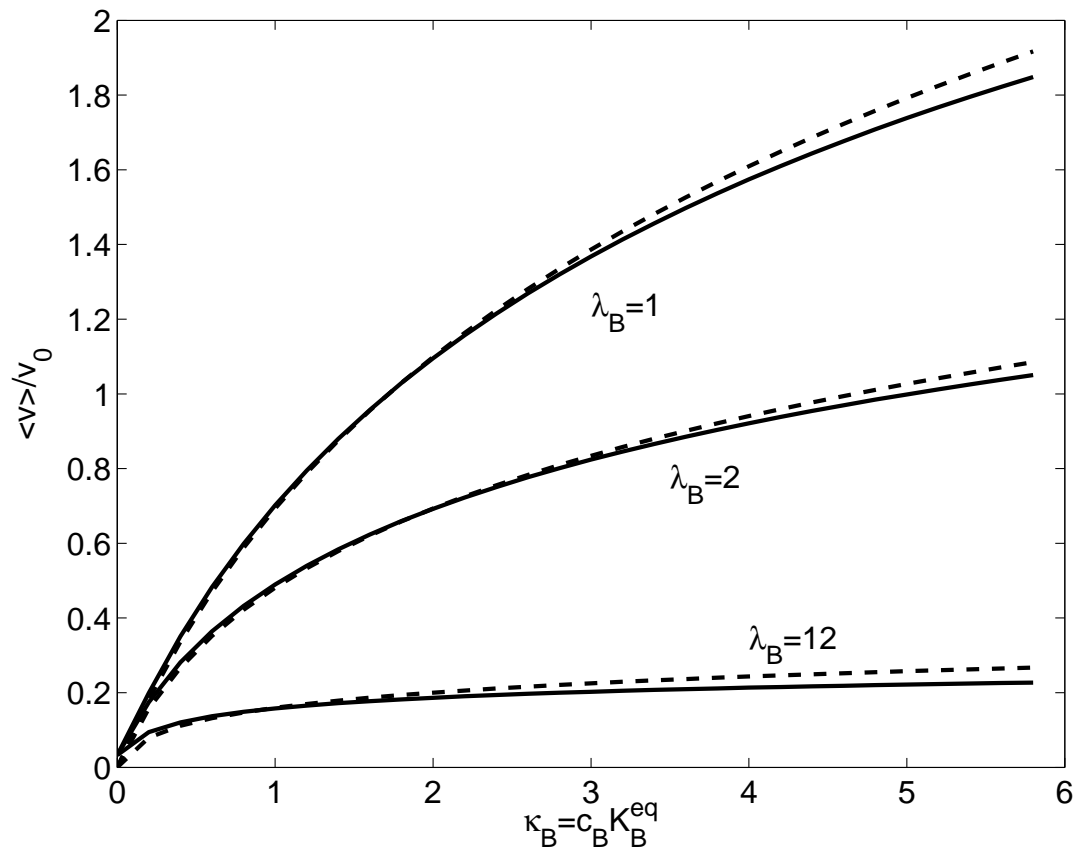}}
  \end{center}
  %\vspace{1cm}
  \caption{The mean velocity $\langle v \rangle$ for translocation of
  a finite sized polymer as a function of binding strength
  $\kappa_B=c_BK_B^{\rm eq}$, for different relative chaperone sizes
  $\lambda_B$ ($c_B$ is the concentration of chaperones, and $K_B^{\rm
  eq}$ is the equilibrium binding constant). The solid lines
  corresponds to the mean velocities as calculated using equations
  \eref{eq:tau}, \eref{eq:v_mean} and the exact expression for the
  force as contained in equation \eref{eq:F_B_general}. The dashed
  curve in connection with the $\lambda_B=1$ line, is the approximate
  result given in equations \eref{eq:v_mean2} together with equation
  \eref{eq:g_B}. The dashed curve with the $\lambda_B=2$ line is the
  approximate expression given in equation \eref{eq:v_mean2} and
  \eref{eq:F_dimer}. For $\lambda_B=12$ the dashed line corresponds to
  the result obtained through equations \eref{eq:algebraic_eq} and
  \eref{eq:F_DiCera}. No binding particles were present on side
  $A$. The effective length of the polymer was taken to be $M=60$. The
  mean velocity increases monotonically with increasing binding
  strength $\kappa_B$. Notice that the deviations between the exact
  (solid curves) and the approximate results (dashed curves) are more
  pronounced for large values of $\kappa_B$.}
  \label{fig:v_mean}
\end{figure}

The mean velocity is obtained on the basis of the force obtained in
the previous subsections together with equations \eref{eq:tau} and
\eref{eq:v_mean}. For the general case (finite sized polymers) the
force expressions as contained in equations \eref{eq:F_A_general} and
\eref{eq:F_B_general} must be used. For sufficiently long polymers we
can ignore the finite-size effect and use equation
\eref{eq:v_mean_long} together with the thermodynamic expressions for
the force derived in the previous subsections. We point out that
$\langle v \rangle$ then (for non-cooperative binding) depends on four
dimensionless variables: $\kappa_A=c_A K_A^{\rm eq}$, $\kappa_B=c_B
K_B^{\rm eq}$, $\lambda_A$, $\lambda_B$ ($c_\gamma$ is the
concentration of chaperones on side $\gamma$ and $K_\gamma^{\rm eq}$
is the equilibrium binding constant for the chaperones on side
$\gamma$, $\gamma$=$A$ or $B$). We can therefore for sufficiently long
polymers write the mean velocity according to
  \be
\langle v \rangle=v_0\left( \frac{|\bar{F}_B|}{F_0}- \frac{|\bar{F}_A|}{F_0}\right)\label{eq:v_mean2}
  \ee
with $v_0=D/\sigma$ and $F_0=k_B T/\sigma$. The relevant expressions
for the forces $\bar{F}_A$ and $\bar{F}_B$ were given in the previous
section: (i) For the general case the forces follow equation
\eref{eq:F_DiCera} and the problem is that of determining the largest
root $\Lambda^{\rm max}$ to the algebraic equation
\eref{eq:algebraic_eq}. (ii) For {\em univalent} binding
($\lambda_\gamma=1$) we use the force expressions according to equation
\eref{eq:F_univalent_coop}. For the case of no cooperativity,
$\omega_\gamma=1$, this equation reduces to the simple result given in
equation \eref{eq:g_B}. (iii) For {\em divalent} binding
($\lambda_\gamma=2$) and non-cooperative interactions ($\w_\gamma=1$)
the forces are given by equation \eref{eq:F_dimer}.  When the
chaperone baths on the two sides contain chaperones of identical
binding strengths $\kappa_A=\kappa_B$ and sizes $\lambda_A=\lambda_B$
the mean velocity is zero, as it should.  In general, however, the
size of the chaperones on side $A$ and side $B$ may differ
$\lambda_A\neq \lambda_B$, which may lead to interesting behaviour of
the mean velocity as a function of $\kappa_A$ and $\kappa_B$ (i.e., of
the chaperone concentration on the two sides). In particular we notice
that the dependence on $\kappa_\gamma$ differs between the cases of
univalent and divalent binding (see equations \eref{eq:g_B} and
\eref{eq:F_dimer}) for non-cooperative binding. The binding strength
$\kappa_\gamma$ is proportional to the concentration of chaperones on
the two sides. Thus by measuring the mean velocity as a function of
concentration of chaperones, it should be possible to reveal the
nature of the binding on the two sides (i.e., the values of
$\lambda_\gamma$ and $\w_\gamma$). In figure \ref{fig:v_mean} we have
plotted $\langle v \rangle $ as a function of $\kappa_B$ for different
$\lambda_B$, assuming no chaperones to be present on side $A$ for
simplicity. The solid lines in figure \ref{fig:v_mean} correspond to
the mean velocity as calculated using equations \eref{eq:tau},
\eref{eq:v_mean} and the exact expression for the force equation
\eref{eq:F_B_general}. The dotted lines corresponds to the approximate
results obtained from equations \eref{eq:v_mean2}, \eref{eq:g_B},
\eref{eq:F_dimer} and \eref{eq:F_DiCera}.  We notice that the
deviation between the above approximate results and the exact result
for the mean velocity is larger for larger values of
$\kappa_\gamma$. This originates from the fact that for large
$\kappa_\gamma$ the finite size correction to the force is more pronounced (see
equation \eref{eq:triangle2}). For finite sized polymers the finite
size correction to the force found here thus a non-negligible plays a role in the
translocation dynamics. However as we increase the length of the
polymer the approximate results (dashed lines) as given above,
coincide with the exact result (solid lines).

A few words on the experimental relevance of the finite size effects
 are in order.  From figure \ref{fig:v_x} we notice that finite size
 effects are prominent for the {\em force} for $m$-values up to $m <
 (3-4)\lambda_B$ for the $\kappa_B$-value chosen in the figure; thus
 typically the larger the chaperones (larger $\lambda_B$) the more
 pronounced is the finite size effect for a given polymer length. We
 note here that it might be possible to measure the finite size effect
 of the force directly; for instance by attaching a bead at one end of
 the polymer and trapping the bead in an optical tweezer, one might
 directly probe the force on the polymer due to the presence of
 chaperones, compare to the experimental setup in reference
 \cite{Salman_etal}. The force is, however, not the usual experimental
 observable. Instead, what is usually obtained in experiments is the
 mean velocity (or mean translocation time). As seen in figure
 \ref{fig:v_mean} the finite size effects are effective also for
 ``long'' polymers ($M >(3-4)\lambda_B$). This is due to the fact that
 the mean velocity, also for long polymers, contains information about
 the dynamics in the small $m$-regime.

%%%%%%%%%%%%%%%%%%% Figure 3 %%%%%%%%%%%%%%%%%%%%%%%%%%%%%%%%

%%%%%%%%%%%%%%%%%%%%%%%%%%%%%%%%%%%%%%%%%%%%%%%%%%%%%%%5

\section{Comparison to electric field induced translocation}\label{sec:comparison}

In this section we compare the binding assisted translocation to
electric field induced translocation. As we have seen in the previous
sections the characteristic velocity due to binding of chaperones
along a polymer is $v_0=D/\sigma$, where $D$ is the polymer diffusion
constant and $\sigma$ is the distance between binding sites. We now
compare $v_0$ to the velocity due to the the presence of an
electrostatic voltage $\triangle V$ across the membrane. If we denote
the linear charge density of the polymer (charge per unit length) by
$\rho$, the electric force on the translocating polymer is $F_{\rm
elec}=\rho \triangle V$, and hence the velocity is
  \be
v_{\rm elec}=\frac{F_{\rm elec}}{\xi}=\frac{D\rho \triangle V}{k_B T},
  \ee
where $\xi=k_B T/D$ is the friction constant for the polymer.  Setting
$v_0=v_{\rm elec}$ we find that we need a voltage across the membrane
  \be
\triangle V=\triangle V_t=\frac{k_B T}{\rho \sigma}
  \ee
in order to get a velocity from the electric field which equals the
velocity due to the chaperones. Let us estimate the voltage for a
(highly) negatively charged polymer like DNA. We then take $\rho\simeq
5$ unit charges/nm, $k_B T\simeq $ 26 meV (room temperature) and
$\sigma\simeq 1$ nm, which gives $\triangle V_t\simeq 5$ mV.  The
typical (resting) potential across the eukaryotic cell membranes is
$\simeq 70$ mV. Thus for charged polymers like DNA it is
``preferable'' to use electric fields for efficient transport. In
contrast, the linear charge density of a protein is sensitive to the
aminoacid sequence and the pH of the solution; at high (low) pH a
protein is typically negatively (positively) charged. Therefore, {\em
non-specific} protein transport cannot in general rely on electric
field induced translocation; this may explain in parts  why nature has invented
the chaperone assisting machinery.

\section{Summary and outlook}\label{sec:summary_outlook}

We have in this work investigated the translocation of a stiff polymer
through a nanopore in a membrane, in the presence of binding particles
(chaperones) that bind to the polymer on both sides of the
membrane. Assuming that the diffusion of chaperones is fast compared
to the rate of translocation we described the process by a
one-dimensional master equation. We closely investigated the
translocation dynamics for the case of reversible binding to the
polymer and found that the dynamics depend on whether the chaperones
bind univalently or multivalently to the polymer. For the case of
univalent binding we derived an analytic finite size correction to the
force exerted on the polymer by the chaperones. In general, the finite
size corrections we quantified in this study may be used to extract
information on the nature of the chaperones from experimental
data. For long polymers a simple expression for the mean velocity of
the polymer through the pore was found. We also discussed the problem
of irreversible binding to the translocating polymer, as well as
compared the effectiveness of binding assisted translocation to
electric field driven translocation.

We want to point out that the case of perfect thermal ratchet translocation
\cite{Simon_Peskin_Oster} (immediate irreversible binding) {\em
cannot} be obtained by simply taking the effective binding strengths to be
infinite in the results in this study. As discussed in
section \ref{sec:general_framework}, for irreversible binding the
chaperones do not have time to unbind during the translocation,
rendering a thermodynamic evaluation (as we have done here) of the
force inapplicable. However, we found that our master equation
approach allows us to {\em formally} describe the case of thermal
ratcheting. It will be interesting to see whether it is possible to
develop a theory that covers both the reversible and irreversible
binding regimes (i.e., arbitrary values of the binding
strength). Possibly techniques and results from the class of the
parking lot models \cite{Krapivsky_BenNaim,Luthi_etal,Talbot_etal}
could prove useful.

We have in this investigation not included entropic effects due to
polymer flexibility.  As noted in the main text, provided that the
chaperone binding does not depend on the curvature of the flexible
polymer the binding force as calculated here and the entropic force
are additive. For not too long flexible polymers the entropic effect
could thus be included in a standard fashion (see e.g.,
\cite{Sung_Park}), but typically these effects in the presence of the
chaperone-generated drift will be negligible for systems relevant to
this study \cite{Lubensky_Nelson}. For very long polymers the dynamics
changes qualitatively, and has to be modelled by a dynamical equation
with memory \cite{Chuang_Kantor_Kardar}. We have also neglected the
volume of the translocating polymer, which should be a fair assumption
for the relevant biological systems.

We have throughout the study assumed that the binding energy for the
chaperones is the same along the polymer. However proteins, RNA and
DNA in general consist of heterogeneous sequences of aminoacids, bases
or base-pairs respectively. It would therefore be interesting to
investigate how heterogeneity in the binding energies along the
polymer affects the translocation dynamics.

It has been suggested that in order for a protein to be able to
translocate it has to be unfolded on the entrance side
\cite{Simon_Peskin_Oster}. The unfolding of a protein in general
requires the presence of chaperones on the entrance side; as we have
seen in this study the presence of such proteins always give an
opposing force compared to the translocation direction. Hence
efficient translocation requires that the amount of binding proteins
on the entrance side is large enough to allow unfolding, but small
enough not to cause a too large opposing force. It will be interesting
to see whether such an optimization concerning the concentration of
``unfolders'' is indeed used in nature. This situation may be improved
by additional protein channels for the chaperones, by which the
relative concentration on both sides of the membrane may be actively
regulated. The possibility also exists that the translocation could be
driven by the refolding of the protein on the target side
\cite{Simon_Peskin_Oster}. Alternatively, in cases where the protein
is synthesised on the entrance side a built-in additional sequence can
inhibit folding. This folding-preventing sequence then has to be
removed on the exit side and then the folding process may assist the
translocation. We note that effects similar to protein folding can
occur for RNA and single stranded DNA in the form of secondary
structure. In principle protein translocation could occur even for an
unfolded protein, provided that the chemical or electric bias is
strong enough. The translocation dynamics could in such case provide
local information about the protein structure, which could find
biotechnological applications, similar to RNA translocation
\cite{Gerland_etal_2003}.

\ack
We would like to thank John Kasianowicz and Ed Di Marzio for interesting
discussions.

\appendix 

\section{Mean translocation time for a constant force}\label{sec:mean_translocation_time}

Let us consider the mean translocation time, as given by equation \eref{eq:tau}. For the case of a constant force $F(m)=F$
the transfer coefficients, equation \eref{eq:t_m}, become
$t^+(m)=(1/\tau_0)(1+F/2F_0)=t^+$ and
$t^-(m)=(1/\tau_0)(1-F/2F_0)=t^-$. \Eref{eq:tau} then becomes:
  \be
\tau = \frac{1}{\tau^+} \sum_{y=0}^M \left(\frac{t^-}{t^+}\right)^y \sum_{z=0}^y \left(\frac{t^+}{t^-}\right)^z.
  \ee
If we now assume that $t^-<t^+$ and use the result for a geometric
series (valid for $Q\neq 1$) $\sum_{z=0}^y Q^z = (1-Q^{y+1})/(1-Q)$ we
find the following expression for the mean translocation time
  \be
\tau=\frac{M}{t^+-t^-}-\frac{t^+}{(t^+-t^-)^2} \left\{1- \left(\frac{t^-}{t^+}\right)^{M+1} \right\} \rightarrow \frac{M}{t^+-t^-}= \frac{M F_0}{F}\label{eq:tau_constant_force},
  \ee
where we in the last step have assumed that $M\gg 1$. For large $M$
(and a constant force), the mean translocation time is thus inversely
proportional to the force, as it should.

\section{Binding partition function}\label{sec:binding_partition_function}

In this appendix we obtain the binding statistical weight
$Z_\gamma(m,n_\gamma)$.

In order to obtain the force on the polymer we must have explicit
expressions for the statistical weights $Z_\gamma(m,n_\gamma)$ (see
equation \eref{eq:F_gamma_def}). This quantity is obtained by
calculating the statistically averaged number of ways to attach
$n_\gamma$ particles on side $\gamma$, divided by a reference
statistical weight $Z_\gamma^{\rm ref}$. We choose $Z_\gamma^{\rm
ref}$ as the number of states in the absence of the polymer (notice
that the choice of $Z_\gamma^{\rm ref}$ is arbitrary, since it
vanishes in equation \eref{eq:F_gamma_def}). Let us first calculate
$Z_\gamma^{\rm ref}$. Denote by $V_A$ the volume of compartment $A$
and by $V_B$ the volume of compartment $B$. Similarly, we assume a
chaperone on side $A$ (side $B$) to occupy a volume $v_{0A}$
($v_{0B}$). There are hence $N_A^{\rm tot}=V_A/v_{0A}$ number of
voxels to put the chaperones on side $A$ (see figure
\ref{fig:geometry}), and similarly $N_B^{\rm tot}=V_B/v_{0B}$ number
of voxels to put the chaperones on side $B$. If we furthermore assume
that there are $N_A$ ($N_B$) chaperones on side $A$ (side $B$), the
number of ways of arranging these particles on the two sides are
  \be
Z_\gamma^{\rm ref}={N_\gamma^{\rm tot} \choose N_\gamma}=\frac{(N_\gamma^{\rm tot})!}{N_\gamma! (N_\gamma^{\rm tot}-N_\gamma)!}\label{eq:Z_ref},
  \ee
which then determine the reference statistical weight. Let us proceed
by calculating the statistical weights $Z_\gamma (m,n_\gamma)$ in the
{\em presence} of the polymer, neglecting the reduction of compartment
volume due to the translocating polymer. As before, on side $B$ the
polymer is divided into $m$ segments such that $m=x/\sigma$, where $x$
is the protrusion distance on side $B$ and $\sigma$ is the size of a
binding site. Similarly on side $A$ there are $M-m=(L-x)/\sigma$
segments, where $M=L/\sigma$ is the total number of binding sites ($L$
is the length of the polymer). We assume that the chaperones cover an
integer $\lambda_\gamma$ ($\ge 1$) number of binding sites if bound to
the polymer on side $\gamma$. The total binding energy for an attached
chaperone is denoted by $\epsilon_\gamma$ ($<0$). The maximum number of
particles that can attach to the polymer on side $B$ is then $n_B^{\rm
max}=[m/\lambda_B]$, i.e., it is the largest integer smaller than or
equal to $m/\lambda_B$. Similarly for side $A$ the maximum number of
attached chaperones is $n_A^{\rm max}=[(M-m)/\lambda_A]$. Denote by
$\Omega_B^{\rm bind}(m,n_B)$ the number of ways of arranging $n_B$
particles onto the $m$ binding sites on side $B$ ($\Omega_B^{\rm
bind}(m,n_\gamma)$ is explicitly given in the main text).  To obtain
the binding statistical weight for side $B$ we have to multiply
$\Omega_B^{\rm bind}(m,n_B)$ by the Boltzmann weight associated with
binding, i.e.,
  \be
Z_\gamma^{\rm bind}(m,n_\gamma)=\Omega_\gamma^{\rm bind}(m,n_\gamma)\exp (-\beta \epsilon_\gamma n_\gamma).\label{eq:Z_bind_B}
  \ee
In order to obtain the full statistical weight for side $B$, we also have to
 account for the fact that when $n_\gamma$ number of particles are
bound to the polymer there are only $N_\gamma-n_\gamma$ numbers of
molecules left in the ``gas'' surrounding the polymer. The number of
states for the ``gas'' (compare equation \eref{eq:Z_ref}) is 
  \be
Z^{\rm
gas}_\gamma(n_\gamma)=\{(N_\gamma^{\rm tot})!\}/\{(N_\gamma-n_\gamma)!
(N_\gamma^{\rm tot}-(N_\gamma-n_\gamma))!\}. 
  \ee
For large $N_\gamma$ (see equation 6.1.47 in reference
\cite{Abramowitz_Stegun}) we have the identity
$(N+a)!/(N+b)!=\Gamma(N+a+1)/\Gamma(N+b+1)\approx
N^{a-b}(1+(a-b)(a+b+1)/(2N)+..)$, where $\Gamma(z)$ is the
$\Gamma$-function. Applying this result, we find that for the case
when a large number of chaperones are present at the two sides: 
  \be
Z^{\rm gas}_\gamma(n_\gamma)= \Phi_\gamma^{n_\gamma}\label{eq:Z_gas_B},
  \ee
where we have introduced the volume fraction on side $\gamma$ as
$\Phi_\gamma\equiv N_\gamma /(N_\gamma^{\rm tot}-N_\gamma)$. Combining
equations\ \eref{eq:Z_bind_B} and \eref{eq:Z_gas_B} we find that the 
statistical weight for side $\gamma$ is:
  \be
Z_\gamma(m,n_\gamma)=Z^{\rm gas}_\gamma (n_\gamma) Z_\gamma^{\rm bind}(m,n_\gamma) = \Omega_\gamma^{\rm bind}(m,n_\gamma) \kappa_\gamma^{n_\gamma}\label{eq:Z_B_spec_app},
  \ee
where we have defined an effective binding strength
  \be
\kappa_\gamma\equiv \Phi_\gamma\exp (\beta |\epsilon_\gamma|)\label{eq:kappa_B}.
  \ee 
For the case of {\em dilute}
solutions ($N_\gamma \ll N^{\rm tot}_\gamma$) the effective binding
strength can be written in the common form: 
  \be
\kappa_\gamma = c_\gamma K^{\rm eq}_\gamma  \label{eq:kappa_gamma_app}
  \ee
where $c_\gamma =N_\gamma/V_\gamma$ is the concentration of
chaperones on side $\gamma$ ($\gamma$=$A$ or $B$), and 
   \be
K^{\rm eq}_\gamma = v_{0\gamma} \exp (\beta |\epsilon_\gamma|)\label{eq:eq_binding_constant}
  \ee
is the equilibrium binding constant \cite{Epstein}.  Equation
 \eref{eq:Z_B_spec_app} together with equation
 \eref{eq:kappa_gamma_app} defines the statistical weights for the two
 sides.

\section{Filling fraction}\label{sec:filling_fraction}

In this appendix we consider the filling fraction of chaperones bound
to the translocating polymer for the cases univalent and divalent of
binding, respectively.

Since we assume that the motion of the chaperones is fast
compared to the rate of translocation through the pore, the expected
numbers $\langle n_A \rangle$ and $\langle n_B \rangle$ of bound chaperones on
the two sides for a given $m$ are well-defined quantities, which are
simply obtained by calculating the expectation value with respect to
the equilibrium distribution, i.e., 
  \be
\langle n_\gamma \rangle =\sum_{n_\gamma=0}^{n_\gamma^{\rm max}}  n_\gamma \rho^{\rm eq}_\gamma(m,n_\gamma) =\frac{\partial }{\partial
\ln \kappa_\gamma } \left( \ln Z_\gamma(m) \right) \label{eq:n_exp_def},
  \ee
where $\gamma$=$A$ or $B$ and the total expected number of bound
chaperones is $\langle n \rangle =\langle n_A \rangle +\langle n_B
\rangle$.

Let us calculate the expected number of bound particles on the two
sides for the case of univalent $\lambda_\gamma=1$ and non-cooperative
binding $\w_\gamma=1$. Using equations
\eref{eq:Z_x_spec} and \ref{eq:n_exp_def} we find $\langle n_\gamma
\rangle = n_\gamma^{\rm max}f_\gamma $ where the filling fractions are
  \be
f_\gamma=\frac{\kappa_\gamma}{1+\kappa_\gamma}=1-(1+\kappa_\gamma)^{-1}\label{eq:f_gamma}.
  \ee
This finding is a standard result for univalent, non-cooperative
binding to a polymer \cite{Wyman_Gill}. We notice that $0\le f_\gamma
\le 1$, and that the chain becomes fully occupied,
$f_\gamma\rightarrow 1$, if the binding strength is very large
$\kappa_\gamma\rightarrow \infty$. When the binding strength is zero
$\kappa_\gamma\rightarrow 0$ there are no chaperones bound, $f_\gamma
\rightarrow 0$, as it should. For positive binding energies
$\epsilon_\gamma\ge 0$ (repulsion) the filling fraction loses its
meaning.  Since for univalent binding (and no cooperativity) the
binding sites are independent, the equilibrium probability $P^{\rm
eq}_{\rm occ}$ that a binding site is occupied equals the filling
fraction, i.e. $P^{\rm eq}_{\rm occ}=f_\gamma$.

We can also calculate $f_\gamma$ for the case of divalent binding
$\lambda_\gamma=2$, and large protrusion distances. Using the results
from subsection \ref{sec:general_results_large_m} we find that
  \be
f_\gamma\approx 1-(1+4\kappa_\gamma)^{-1/2}\label{eq:f_gamma_dimer},
  \ee
for divalent binding. We have that $0\le f_\gamma \le 1$,
$f\rightarrow 1$ for $\kappa_\gamma\rightarrow \infty$, and
$f_\gamma\rightarrow 0$ for $\kappa_\gamma\rightarrow 0$ as it
should. We notice that for divalent binding the polymer reaches its
fully occupied state $f_\gamma =1$ ``slower'' with $\kappa_\gamma$
than for the case of univalent binding (see equation \eref{eq:f_gamma}).
\footnote[1]{Equations \eref{eq:f_gamma} and \eref{eq:f_gamma_dimer} also follows
straightforwardly from the McGhee-von Hippel binding isotherm
\cite{McGhee_vonHippel}:
  \bes
\frac{f_\gamma}{\lambda_\gamma}=\kappa_\gamma (1-f_\gamma)\left( \frac{1-f_\gamma }{1-(\lambda_\gamma-1)f_\gamma/\lambda_\gamma}\right)^{\lambda_\gamma -1}
  \ees
which expresses the fraction $f_\gamma$ of occupied binding sites as a
function of $\kappa_\gamma$ and $\lambda_\gamma$ for binding to an
infinitely long polymer. From the above expression we notice that for
large chaperones $\lambda_\gamma\gg 1$ the filling fraction takes the
simple form $f_\gamma=1-(1+\lambda_\gamma\kappa_\gamma)^{-1}$.} This
is intuitively clear, as for divalent binding those configurations have
to be overcome in which vacant spots of the size of one binding site
have to disappear in order to reach $f_\gamma=1$.

%%%%%%%%%%%%%%% Bibliography %%%%%%%%%%%%%%%%%%%%

%%%%%%%%%%%%%%%%%% Figures %%%%%%%%%%%%%%%%%%%

\end{document}